\documentclass[twocolumn,amsmath,amssymb]{revtex4}

\draft 

\usepackage{graphicx}
\usepackage{dcolumn}
\usepackage{bm}

\begin{document}

\preprint{AIP/123-QED}

\title{A circuit analysis of an \emph{in situ} tunable radio-frequency quantum point contact}

\author{T. M\"uller}
\email{thommuel@phys.ethz.ch}\altaffiliation[Present address:
]{Department of Chemistry, University of Cambridge, Cambridge CB2
1EW, UK}
\author{T. Choi}
\author{S. Hellm\"uller}
\author{K. Ensslin}
\author{T. Ihn}
\affiliation{Solid State Physics Laboratory, ETH Z\"urich, 8093
Z\"urich, Switzerland}
\author{S. Sch\"on}
\affiliation{FIRST Laboratory, ETH Z\"urich, 8093
Z\"urich, Switzerland}

\date{\today}

\begin{abstract}
A detailed analysis of the tunability of a radio-frequency quantum
point contact setup using a $C-LCR$ circuit is presented. We
calculate how the series capacitance influences resonance frequency
and charge-detector resistance for which matching is achieved as
well as the voltage and power delivered to the load. Furthermore, we
compute the noise contributions in the system and compare our
findings with measurements taken with an etched quantum point
contact. While our considerations mostly focus on our specific
choice of matching circuit, the discussion of the influence of
source-to-load power transfer on the signal-to-noise ratio is valid
generally.
\end{abstract}

\maketitle

\section{Introduction\label{sec:intro}}

\begin{figure}[htb]
        \centering
\includegraphics{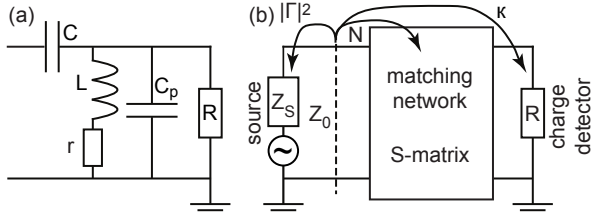}
\caption{\small (a) Simplified diagram of the resonant circuit used
in our experiments, including a resistance $r$ in series to the
inductor to account for losses. (b) Schematics of a general two-port
matching network in which a fraction $|\Gamma|^2$ of the incoming
power is reflected, a fraction $\kappa$ is delivered to the load
(charge detector), and a fraction $N$ is dissipated in the matching
circuit. Part (b) of this figure is adapted from
Ref.~\onlinecite{roschier:04}.}\label{Fig:SimpleCircuit}
\end{figure}

Understanding the electrical behavior of a given resonant circuit is
key for maximizing the charge sensitivity of radio-frequency (rf)
reflectometry experiments such as rf single-electron
transistor\cite{lu:03} or quantum point
contact\cite{mueller:07,reilly:07,cassidy:07} charge sensing. In
view of this goal, Roschier \emph{et al.}\cite{roschier:04} have
presented a thorough analysis of the most commonly used $L-CR$
matching network \cite{schoelkopf:98}. In this work, we pursue a
rather similar purpose but focus on the $C-LCR$
circuit\cite{sillanpaa:04,sillanpaa:05} shown in
Fig.~\ref{Fig:SimpleCircuit}(a) which can be adapted for \emph{in
situ} tunability\cite{mueller:10}. As we highly appreciate the
detailedness of Ref.~\onlinecite{roschier:04} we shall try to keep
up a similar level of depth while following a complementary path and
highlighting different aspects.

We will start with a general observation on the influence of the
fraction of available power actually delivered to the matched load
on the system's charge sensitivity in Sec.~\ref{sec:transfer}.
Subsequently, we will detail our choice of tunable matching circuit
in Sec.~\ref{sec:tuning}, whereafter our discussion will become
specific to this type of circuit. Moreover we will calculate the
maximally applicable voltage given a certain allowed voltage drop
over the load and the corresponding input power. We will use these
results to determine the network's anticipated reaction to a change
in the system's charge state. In Sec.~\ref{sec:noise} we will
compute the relevant noise contributions and compare their magnitude
to the charge-response in order to predict the detector's charge
sensitivity. Finally, we will verify our findings with measurements
taken using a quantum point contact etched into the two-dimensional
electron gas of a GaAs/AlGaAs heterostructure in
Sec.~\ref{sec:measurement}.

Since our research is directed towards quantum point contact (QPC)
charge-detection on semiconductor quantum dots we assume the load to
have a linear $I-V$ curve which facilitates the analysis compared to
the use of a differential resistance\cite{korotkov:99}. In real
experimental life, however, this assumption may sometimes be
inaccurate even at low bias as the exact potential landscape can
drastically affect the $I-V$ characteristics.

\section{Power Transfer\label{sec:transfer}}

In this first section we will deduce a general analytic expression
for the signal-to-noise ratio (SNR) - inversely proportional to the
square of the charge sensitivity - obtained with rf reflectometry
measurements. To this end we have to compute the magnitude of the
signal - given by the change in reflection coefficient upon addition
of a single charge to the system times the input voltage - and the
total noise voltage or power.

If any amount of power is applied to a matching network, only a
certain fraction $\kappa$ of the power supplied by the source is
delivered to the load while another fraction $N$ is dissipated by
the lossy circuit and a part $|\Gamma|^2$ is reflected. This
situation is sketched in Fig.~\ref{Fig:SimpleCircuit}(b) which has
been adapted from Ref.~\onlinecite{roschier:04}. There is a source
with impedance $Z_S$ connected to a transmission line with impedance
$Z_0$ leading to the matching network with $S$-matrix $S$ directly
connected to a load $R$. While the load may have a complex impedance
which can be utilized for instance in dispersive
readout\cite{petersson:10}, we restrict ourselves to real values of
$R$. Nevertheless, generalization is readily obtained. Note that in
reflectometry experiments the reflected wave is measured through an
amplifier which, for simplicity, we assume to be matched to $Z_0$.
Naturally, the fractions introduced above follow the equation
\begin{equation}\label{eq:fractions}
    1=|\Gamma|^2+\kappa+N.
\end{equation}
As shown for instance in the chapter on microwave amplifier design
in Ref.~\onlinecite{pozar:11} and discussed in
Ref.~\onlinecite{roschier:04}
\begin{eqnarray}
  \label{Eq:GammaPozar} \Gamma &=& \frac{Z-Z_0}{Z+Z_0}=S_{11}+\frac{S_{12}S_{21}\Gamma_R}{1-S_{22}\Gamma_R}\qquad\textrm{and} \\
  \label{Eq:KPozar} \kappa &=&
  \frac{\left|S_{21}\right|^2\left(1-\left|\Gamma_S\right|^2\right)\left(1-\left|\Gamma_R\right|^2\right)}{\left|1-S_{22}\Gamma_R\right|^2\left|1-\Gamma_S\Gamma\right|^2},
\end{eqnarray}
where $S_{ij}$ are the matching network's $S$-matrix elements and
\begin{eqnarray}
  \label{Eq:GammaRPozar} \Gamma_R &=& \frac{R-Z_0}{R+Z_0}, \\
  \label{Eq:GammaSPozar} \Gamma_S &=& \frac{Z_S-Z_0}{Z_S+Z_0}.
\end{eqnarray}
The characteristic impedance of the rf lines $Z_0$ is usually
$50~\Omega$. For a matched source (which we will assume for the
remainder of this work), $\Gamma_S$ is equal to zero, and
Eq.~(\ref{Eq:KPozar}) simplifies to
\begin{equation}\label{Eq:KSimplePozar}
   \kappa = \frac{\left|S_{21}\right|^2\left(1-\left|\Gamma_R\right|^2\right)}{\left|1-S_{22}\Gamma_R\right|^2}.
\end{equation}

To go beyond this we will determine a generally valid expression for
the reflection coefficient's sensitivity on $R$. Therefore, we
calculate the absolute value of the differential change in $\Gamma$
via
\begin{equation}\label{Eq:DeltaGammaDiff}
    \left|\Delta\Gamma\right|\simeq\left|\frac{\partial \Gamma}{\partial R}\Delta R\right|=\left|\frac{\partial \Gamma}{\partial \Gamma_R}\right|\left|\frac{\partial \Gamma_R}{\partial R}\right|\left|\Delta
    R\right|,
\end{equation}
as measured using in-phase and quadrature amplitudes in homodyne
detection. Similarly, one could study the phase change or the change
in absolute value of the reflection coefficient, for instance when
relying on heterodyne detection.

The first part of Eq.~(\ref{Eq:DeltaGammaDiff}) can be evaluated as
\begin{eqnarray}
\nonumber    \left|\frac{\partial \Gamma}{\partial \Gamma_R}\right| & = &\left|\frac{(1-S_{22}\Gamma_R)S_{12}^2-S_{12}^2\Gamma_R\times(-S_{22})}{(1-S_{22}\Gamma_R)^2}\right| \\
\label{Eq:DiffGammaR}    & = &
\left|\frac{S_{12}^2}{(1-S_{22}\Gamma_R)^2}\right|=\frac{\kappa}{1-\left|\Gamma_R\right|^2},
\end{eqnarray}
using $S_{12}=S_{21}$ in reciprocal two-port networks. The second
part of Eq.~(\ref{Eq:DeltaGammaDiff}) is
\begin{equation}\label{Eq:DiffGammaBR}
    \left|\frac{\partial \Gamma_R}{\partial
    R}\right|=\frac{2Z_0}{(R+Z_0)^2}.
\end{equation}
Combining Eqs.~(\ref{Eq:DeltaGammaDiff}), (\ref{Eq:DiffGammaR}), and
(\ref{Eq:DiffGammaBR}) thus yields
\begin{equation}\label{Eq:DiffGammaTotal}
    \left|\Delta \Gamma\right|\simeq\left|\frac{\kappa}{1-\left|\Gamma_R\right|^2}\frac{2Z_0}{(R+Z_0)^2}\Delta R\right|=\frac{\kappa|\Delta R|}{2R}.
\end{equation}

We want to stress that this analysis is valid for any reciprocal
linear two-port network, independent of the specifics of $S_{ij}$,
as long as $\Gamma$ can be linearized in $R$ over the range $\Delta
R$. It is important to bear in mind, though, that $\kappa$ strongly
depends on the load (charge detector) resistance $R$ and
consequently, the largest change in reflection coefficient is not
necessarily observed at maximum $\Delta R/R=\Delta G/G$.

\begin{figure}[htb]
        \centering
\includegraphics{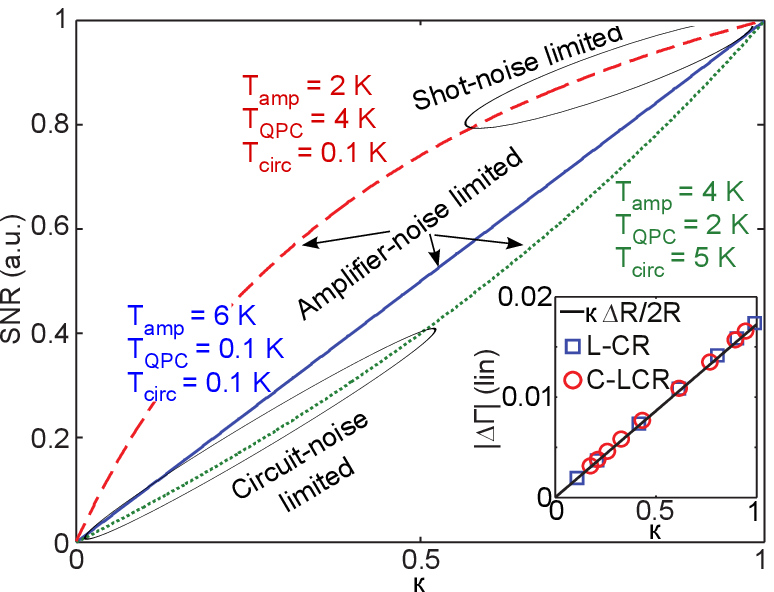}
\caption{\small Relative signal-to-noise ratio as a function of
power-transfer coefficient $\kappa$. The plots represent
Eq.~(\ref{Eq:SNRkappaTemp}) with the parameters
$T_\textrm{amp}=6~\textrm{K}$, $T_\textrm{QPC}=0.1~\textrm{K}$,
$T_\textrm{circ}=0.1~\textrm{K}$ (solid blue),
$T_\textrm{amp}=2~\textrm{K}$, $T_\textrm{QPC}=4~\textrm{K}$,
$T_\textrm{circ}=0.1~\textrm{K}$ (dashed red), and
$T_\textrm{amp}=4~\textrm{K}$, $T_\textrm{QPC}=2~\textrm{K}$,
$T_\textrm{circ}=5~\textrm{K}$ (dotted green). Inset: Change in
reflection coefficient at resonance for $R=29$ to
$28~\textrm{k}\Omega$ as a function of $\kappa$ calculated from
Eq.~(\ref{Eq:DiffGammaTotal}) (solid line) and via impedance for a
$L-CR$ (blue squares) and $C-LCR$-circuit (red circles). The power
transfer coefficient $\kappa$ is varied through $r$ (in series to
the parallel capacitance $C$ for $L-CR$ and the inductance $L$ for
$C-LCR$, respectively), keeping the load resistance for which best
matching is achieved at $30~\textrm{k}\Omega$ by changing $C$ (in
both cases) and if necessary $L$ (only in case of the
$L-CR$-circuit) accordingly.}\label{Fig:SNRvsKappa}
\end{figure}

The change $\Delta\Gamma$ of the reflection coefficient as a
function of $\kappa$, given by Eq.~(\ref{Eq:DiffGammaTotal}), is
shown as the black line in the inset of Fig.~\ref{Fig:SNRvsKappa}
for a change in resistance from 29 to $28~\textrm{k}\Omega$,
realistic as a change in QPC resistance upon tunneling of an
electron into or out of a proximal quantum dot. The blue squares
(red circles) are calculations of
$\Delta\Gamma=\Gamma(R_1)-\Gamma(R_2)$ where $\Gamma$ is determined
from the impedance of a $L-CR$ ($C-LCR$) circuit and $\kappa$ is
varied by introducing a parasitic resistance $r$ in series to $C$
($L$). The parameters $C$ - and in case of the $L-CR$ circuit also
$L$ - are chosen individually for every data point such that the
load is matched at $R=30~\textrm{k}\Omega$. Since
Eq.~(\ref{Eq:DiffGammaTotal}) is exact as long as $\Gamma$ can be
linearized over the range $\Delta R$, the perfect agreement between
the direct (blue squares and red circles) and indirect (black line)
calculation is not particularly surprising.

Now that we know the circuit's response, e.g. due to charge
tunneling into a capacitively coupled quantum dot, we can compute
the signal voltage by multiplying $\Delta\Gamma$ with the input
voltage. If we need to limit the voltage drop over the load,
henceforth assumed to be a QPC, to a given value $V_\textrm{QPC}$ -
which is for instance necessary to avoid excessive back-action in
charge-detection measurements - the tolerated input voltage and
power are fixed through\cite{roschier:04}
\begin{eqnarray}\label{Eq:Vin}
    V_\textrm{in}&=&V_\textrm{QPC}\sqrt{\frac{Z_0}{\kappa R}},\\
    \label{Eq:Pin}P_\textrm{in}&\equiv&P_\textrm{0}-P_\textrm{refl}=\frac{\left|V_\textrm{in}\right|^2}{Z_0}\left(1-|\Gamma|^2\right),
\end{eqnarray}
where $V_\textrm{in}$ is the amplitude of the voltage incident from
the source to the $S$-matrix and $V_\textrm{QPC}$ is the voltage
applied to the load, which is the sum of incoming and reflected
voltages, and $P_\textrm{in}$ is the power delivered to the entire
network given by the difference between applied power $P_\textrm{0}$
and reflected power $P_\textrm{refl}$. Hence the signal amplitude,
defined as the change of the amplitude of the wave reflected at the
$S$-matrix, is given by
\begin{equation}\label{Eq:VSignal}
V_\textrm{S}=\left|V_\textrm{in}\times\Delta
\Gamma\right|=V_\textrm{QPC}\sqrt{\frac{Z_0\kappa}{R}}\frac{|\Delta
    R|}{2R}.
\end{equation}
For QPC charge detection, $V_\textrm{QPC}\simeq300~\mu\textrm{V}$
has been found to be a good compromise between the desire to
maximize the signal but minimize back-action\cite{gustavsson:07}.

By assigning an equivalent noise temperature to all relevant noise
sources - QPC ($T_\textrm{QPC}$, thermal and shot noise; typically a
few K), matching network ($T_\textrm{circ}$, thermal noise; mK to a
few K), and low-noise amplifier ($T_\textrm{amp}$; of the order of a
few K for state-of-the-art devices) - we can conveniently express
the overall noise as measured over the matched amplifier input
resistor by
\begin{equation}\label{Eq:SystemNoise}
    T_\Sigma=T_\textrm{amp}+NT_\textrm{circ}+\kappa T_\textrm{QPC}.
\end{equation}
Note that the circuit noise temperature can be determined as the
concerted action of two correlated noise sources at either input of
the two-port network characterizing the matching
cirucit\cite{twiss:55}. Also, we have chosen a simplified version of
the amplifier noise temperature as compared to
Ref.~\onlinecite{roschier:04} where a noise-wave description was
used.

Thus, we can calculate the signal-to-noise ratio at the amplifier
output via
\begin{eqnarray}
    \nonumber\textrm{SNR}&=&|\Delta\Gamma|^2\times\frac{P_\textrm{0}}{P_N}\\
    \label{Eq:SNRkappa}&=&\frac{\left(\Delta R/2R\right)^2\times\kappa V_\textrm{QPC}^2/R}{k_B\left(T_\textrm{amp}+\kappa T_\textrm{QPC}+NT_\textrm{circ}\right)\Delta f}.
\end{eqnarray}
If $R$ is considered to be kept fixed 
we immediately arrive at
\begin{equation}\label{Eq:SNRkappaTemp}
    \textrm{SNR}\propto \frac{1}{T_\textrm{QPC}+\frac{1}{\kappa}T_\textrm{amp}+\frac{1-\kappa}{\kappa}T_\textrm{circ}}.
\end{equation}
It should be emphasized that the SNR \emph{always} increases with
$\kappa$. Clearly, if there is no power transfer between source and
load ($\kappa=0$), the signal-to-noise ratio vanishes, and if all
the power is delivered to the load ($\kappa=1$) the SNR is inversely
proportional to $T_\textrm{amp}+T_\textrm{QPC}$.

The main graph of Fig.~\ref{Fig:SNRvsKappa} visualizes
Eq.~(\ref{Eq:SNRkappaTemp}) where the SNR is computed as a function
of $\kappa$ and normalized by $(T_\textrm{amp}+T_\textrm{QPC})^{-1}$
for three different cases with $
T_\textrm{amp}=2~\textrm{K},~T_\textrm{QPC}=4~\textrm{K},~T_\textrm{circ}=0.1~\textrm{K}$
(dashed red line),
$T_\textrm{amp}=6~\textrm{K},~T_\textrm{QPC}=0.1~\textrm{K},~T_\textrm{circ}=0.1~\textrm{K}$
(solid blue line), and
$T_\textrm{amp}=4~\textrm{K},~T_\textrm{QPC}=2~\textrm{K},~T_\textrm{circ}=5~\textrm{K}$
(dotted green line), respectively\cite{NoteTemp}. In the first case
for large power transfer the noise at the amplifier output is
dominated by the load's noise temperature - realistically QPC shot
noise - while at low power transfer and in the second case the
dominant contribution originates from the amplifier - which is the
case for low QPC bias or comparably large amplifier noise. Note that
the SNR for the case with $T_\textrm{amp}\gg
\{T_\textrm{QPC},T_\textrm{circ}\}$ is an almost straight line with
a slope of roughly $1/T_\textrm{amp}$ and always lies below the
shot-noise limited case if the sum $T_\textrm{amp}+T_\textrm{QPC}$
is equal for both cases. This follows from the suppression of
$T_\textrm{QPC}$ with $\kappa$. If the matching network adds a
considerable amount of thermal noise the SNR is poorer, which can be
particularly painful at low power transfer.

If an experiment is dominated by amplifier noise, we can assume that
the total noise temperature $T_\Sigma$ is independent of the QPC
resistance. Then, since $\kappa$ can be shown to be inversely
proportional to $R$ for $R\gg Z_0$ (see
Eq.~(\ref{Eq:KSimplePozar})), we obtain
\begin{equation}\label{Eq:SNRRinfty}
    \textrm{SNR}_{R\rightarrow\infty}\propto\left(\frac{\Delta R}{R^2}\right)^2\approx(\Delta
    G)^2,
\end{equation}
where the last approximation is valid if $G\gg\Delta G$.

We can also estimate the increase in SNR expected for experiments
carried out in a dilution refrigerator as compared to a
variable-temperature insert at 2 K; for a fixed amplifier noise
temperature of 2 K and reducing $T_\textrm{QPC}$ as well as
$T_\textrm{circ}$ from 2 to $0.1~\textrm{K}$ and assuming
$\kappa\approx1/3$, the signal-to-noise ratio is doubled. And this
does not yet take into account that a nanostructure's resistance
typically becomes more sensitive to electrostatic potentials at
lower temperatures.

From the above considerations it is clear that having low losses in
the matching circuit - no matter which type - is crucial. Potential
strategies to reduce losses are the use of low-loss printed circuit
boards as well as superconducting air or on-chip
inductors\cite{xue:07} and waveguides.

With this we conclude the generally valid part and turn towards our
choice of a tunable $C-LCR$ circuit.

\section{Tuning of the Matching Network\label{sec:tuning}}

To compensate for large stray capacitances in parallel to the load
when using a ceramics chip-carrier design or working with structures
having large back gates such as graphene devices\cite{mueller:12},
we have selected a $C-LCR$ circuit as a matching network. The series
capacitance $C$ can then be set by an \emph{in situ} tunable
varactor diode\cite{mueller:10}. In the following we will
investigate the effect of changing this capacitance on the frequency
response, on matching at the resonance frequency, and on the maximal
voltage and power we are allowed to send into the matching circuit.

\begin{table}
\begin{center}
\begin{tabular}{l l l}
  Variable & Description & Value \\
  \hline\noalign{\smallskip}\hline\noalign{\smallskip}
  $C$ & Tunable series capacitance & 1.5 pF \\
  $C_p$ & Parallel (stray) capacitance & 2.6 pF \\
  $L$ & Parallel inductance & 150 nH \\
  $r$ & Parasitic resistance & 2 or 5 $\Omega$ \\
  $R$ & Load resistance & 50 k$\Omega$ \\
  $T_\textrm{amp}$ & Amplifier noise temperature & 2 K \\
  $T_\textrm{QPC}$ & Load noise temperature & 2 K \\
  $T_\textrm{circ}$ & Circuit noise temperature & 5 K \\
  \hline
\end{tabular}
\parbox{\columnwidth}{\caption{\small Typical values for the
parameters used in this work.}\label{Tab:Values}}
\end{center}
\end{table}

Since we will be using numerous parameters, we define our standard
set of numerical values in Tab.~\ref{Tab:Values} for the reader's
comfort. Nevertheless, we will still continue to note the utilized
values in the main text.

To facilitate our analysis we consider the simplified circuit model
shown in Fig.~\ref{Fig:SimpleCircuit}(a). This circuit assumes all
stray capacitances - which we will combine in the parameter $C_p$ -
to be in parallel to the load and all losses of the matching network
to be in series to the inductor\cite{Note1} (denoted by a resistance
$r$), neglects inductances of all elements including bond
wires\cite{Note2}, and regards the load (QPC) as a purely real
impedance $R$. Comparing the calculated frequency response of this
simplified circuit with a full circuit model including all the
aspects mentioned above reveals that they are negligible indeed (not
shown). The value of $r$, however, must be determined from
experimentally obtained matching curves and amounts to roughly
$5~\Omega$ when using a GaAs QPC etched into a shallow
two-dimensional electron gas (with InAs nanowires deposited on top
of the chip) as a load and an inductance of $150~\textrm{nH}$,
$3.5~\Omega$ for an etched graphene nanoconstriction and
$100~\textrm{nH}$, or $2~\Omega$ when matching an AFM-defined GaAs
QPC (without nanowires) and with an inductance of
$180~\textrm{nH}$\cite{Note3}. At this point we unfortunately cannot
say where this rather large difference is originating from - a quite
detailed investigation of the influence of the sample geometry (such
as performed in Ref.~\onlinecite{hellmueller:12}) and substrate used
would be required for this.

The impedance of this simplified circuit is given by
\begin{equation}\label{Eq:Zsimple_r}
    Z=\frac{-i}{\omega C}+\left(\frac{1}{R}+i\omega C_p+\frac{1}{i\omega L+r}\right)^{-1}.
\end{equation}
If $L$, $C_p$, $r$, and $R$ are assumed to be fixed by the circuit
and the load, the matching conditions
\begin{eqnarray}
  \textrm{Re}(Z) &=& 50~\Omega \\
  \textrm{Im}(Z) &=& 0
\end{eqnarray}
determine $C$ and $\omega$. In particular, a tunable capacitance
thus allows us to achieve perfect matching ($\Gamma=0$) for any $R$
provided that the range of $C$ is large enough.

\begin{figure}[htb]
        \centering
\includegraphics{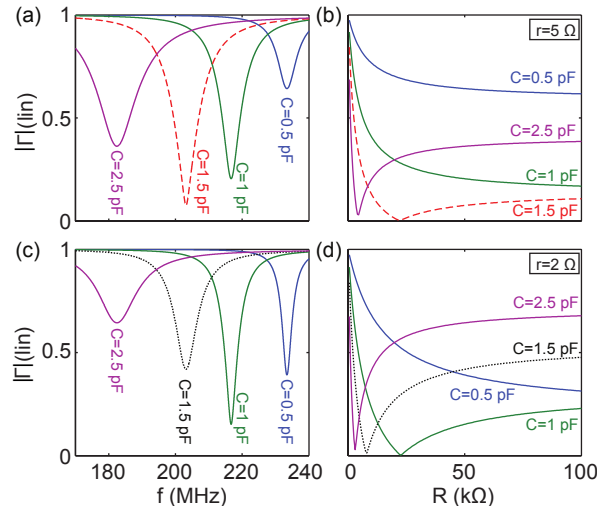}
\caption{\small (a) Influence of the series capacitance on the
frequency response of the matching circuit shown in
Fig.~\ref{Fig:SimpleCircuit}. The parameters used for this
calculation are $C_p=2.6~\textrm{pF}$, $L=150~\textrm{nH}$,
$r=5~\Omega$, $R=50~\textrm{k}\Omega$. The tunable capacitance $C$
is increased from $0.5~\textrm{pF}$ (rightmost blue curve) to
$2.5~\textrm{pF}$ (leftmost purple curve). (b) Reflection
coefficient as a function of QPC resistance at the resonance
frequency. (c) and (d) As in (a) and (b), but with
$r=2~\Omega$.}\label{Fig:Influence}
\end{figure}

Calculating the frequency response of the reflection coefficient
using Eqs.~(\ref{Eq:GammaPozar}) and (\ref{Eq:Zsimple_r}) for
different series capacitances $C$ yields
Figs.~\ref{Fig:Influence}(a) and (c). For these calculations we have
used $R=50~\textrm{k}\Omega$, $C_p=2.6~\textrm{pF}$,
$L=150~\textrm{nH}$, and $r=5~\Omega$ ($r=2~\Omega$). We can clearly
see that an increase in $C$ reduces the resonant frequency, as
expected. Furthermore, both $C$ and $r$ alter the depth of the
resonance. This is more easily visible in
Figs.~\ref{Fig:Influence}(b) and (d), where the reflection
coefficient is calculated as a function of $R$ at the resonance
frequency. The latter is determined by numerically finding the
minimum of $\Gamma$ at $R=50~\textrm{k}\Omega$. Increasing $C$
shifts the resistance $R$ for which optimal matching is achieved to
lower values, while increasing $r$ shifts these values to higher
$R$. If $r$ is too large or if $C$ is too small perfect matching is
never achieved. If $C$ is too large, optimal matching occurs at low
resistances, where charge detectors are generally less sensitive. It
can also be noted that the width of the resonance is decreasing with
increasing resonance frequency.

\begin{figure}[htb]
        \centering
        \includegraphics{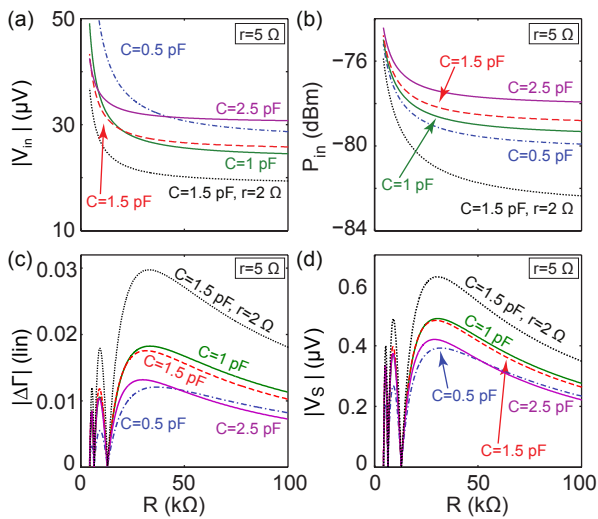}
\caption{\small (a) Absolute value of the maximally allowed rf
voltage at resonance for a given QPC rms voltage of
$300~\mu\textrm{V}$ as a function of $R$ with different serial
capacitance values from $0.5~\textrm{pF}$ to $2.5~\textrm{pF}$. For
the dotted black curve, the parasitic resistance $r$ was chosen to
be $2~\Omega$ instead of $5~\Omega$. (b) Power delivered to the
matching network according to Eq.~(\ref{Eq:Pin}). (c) Absolute value
of the changes in reflection coefficients from
Figs.~\ref{Fig:Influence}(b) and (d) using a change in $R$ modeled
by a QPC in the saddle-point approximation with a change in
potential of $50~\mu\textrm{eV}$ due to an electron entering a
nearby dot. (d) Change in reflected voltage due to electron
tunneling in a nearby dot taking into account the maximally
tolerable input voltage $V_\textrm{in}$ calculated in
Eq.~(\ref{Eq:Vin}) and shown in (a).}\label{Fig:Vrefl}
\end{figure}

Now that we are convinced that we can obtain good matching through
tuning of $C$ and selection of the proper measurement frequency we
want to compute the magnitude of the signal we can expect for a
given change in load resistance (due to the addition of charge on
the quantum dot). Therefore we will first determine numerically the
voltage and power we may apply if the voltage drop over the load
should be maximally $300~\mu\textrm{V}$ (see Eq.~(\ref{Eq:Vin}) in
Sec.~\ref{sec:transfer}). The value of $\kappa$ can be calculated
with Eq.~(\ref{Eq:KSimplePozar}) using
$S$-matrices\cite{pozar:11,roschier:04} or, alternatively, using
$Z$-matrices. Either way leads to the same result, and the maximal
input voltage and the power $P_\textrm{in}$ thus delivered to the
matching network (including load) at resonance are presented in
Figs.~\ref{Fig:Vrefl} (a) and (b) for our usual parameters of
$C_p=2.6~\textrm{pF}$, $L=150~\textrm{nH}$, and $r=5~\Omega$
($r=2~\Omega$ for the dotted black curve). It is visible that worse
matching (see Fig.~\ref{Fig:Influence}) means we can apply a higher
voltage which is reasonable since in this case less signal couples
to the load. For a parasitic resistance of $r=5~\Omega$ the input
voltage is in the range of $30~\mu\textrm{V}$, whereas for
$r=2~\Omega$ we are only allowed to apply $\sim20~\mu\textrm{V}$.
The powers delivered to the circuit are of the order of
$-80~\textrm{dBm}$ which is $10~\textrm{pW}$.

To assess the expected signal height we still need the change in
reflection coefficient upon addition of a single charge to the
system. Towards this end we focus again on QPC charge detection and
model the QPC with a saddle-point potential\cite{buettiker:90} with
$\hbar\omega_x=1~\textrm{meV}$ and $\hbar\omega_y=2~\textrm{meV}$ at
$T=0$. Then, we assume that the addition of a single unit charge to
the charge detector's environment changes its chemical potential by
$50~\mu\textrm{eV}$\cite{Note4}, since such a potential difference
leads to a conductance change of up to $\Delta
G\approx0.08\times2e^2/h$ which is quite realistic for an InAs
nanowire quantum dot self-aligned to a GaAs QPC\cite{shorubalko:08}.
In combination with the reflection coefficients obtained in
Figs.~\ref{Fig:Influence}(b) and (d) this estimated change in QPC
conductance leads to a change in $\Gamma$ as shown in
Fig.~\ref{Fig:Vrefl}(c). The line shape of these curves originates
from the step-like resistance behavior of a QPC. The magnitudes of
the reflection coefficient changes amount from 0.01 for worse
matching up to 0.03 (on a linear scale) for better matching. Since
at superior matching the maximally allowed input voltage is smaller,
the total reflected voltage determined by Eq.~\ref{Eq:VSignal} and
depicted in Fig.~\ref{Fig:Vrefl}(d) depends less strongly on the
quality of the matching, but still better matching and lower losses
lead to a significantly higher signal amplitude. Signal amplitudes
of several hundreds of nanovolts are thus quite realistic for a
highly coupled dot-charge detector system.

We have now found a reliable estimate for the expected signal height
by assessing the change in reflection coefficient due to a modeled
change in QPC resistance and taking into account a maximally allowed
voltage drop over the charge detector. In the next section we will
compare this signal amplitude to the total noise in our system to
predict the signal-to-noise ratio or charge sensitivity of our
charge-detection measurements.

\section{Noise Analysis\label{sec:noise}}

We will start this section by identifying and sizing the relevant
noise sources in our measurement setup. The sum of these noise
contributions will then be checked against the signal height
assessed in the previous section to estimate the signal-to-noise
ratio of a charge detection experiment.

\begin{figure}[htb]
        \centering
        \includegraphics{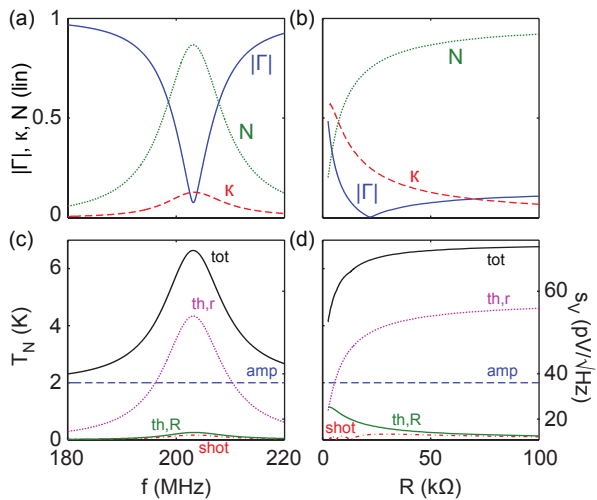}
\caption{\small (a) and (b) Absolute value of the reflection
coefficient $\Gamma$ (solid blue line) and the power-transfer
coefficients $\kappa$ (dashed red) and $N$ (dotted green) as a
function of frequency and QPC resistance at resonance, respectively.
(c) and (d) Frequency and QPC resistance dependence of the
equivalent noise temperatures of all relevant noise sources. These
include QPC shot noise as expected at zero temperature (dash-dotted
red), thermal noise of the QPC at zero bias (solid green), thermal
noise in $r$ (dotted pink), and amplifier noise (dashed blue). The
sum of all noise sources is plotted as a solid black line. The
parameters used are $C=1.5~\textrm{pF}$, $C_p=2.6~\textrm{pF}$,
$L=150~\textrm{nH}$, $r=5~\Omega$, $R=50~\textrm{k}\Omega$,
$f=203.19~\textrm{MHz}$, $T_\textrm{amp}=2~\textrm{K}$,
$T_\textrm{QPC}=2~\textrm{K}$, $T_\textrm{circ}=5~\textrm{K}$,
$V_\textrm{QPC}=300~\mu\textrm{V}$, and $Z_\textrm{amp}=50~\Omega$
for all plots.}\label{Fig:PNnoise}
\end{figure}

The noise power spectral density of the QPC at frequencies much
lower than temperature and voltage bias ($eV,~k_BT\gg\hbar\omega$)
is given by\cite{khlus:87}
\begin{eqnarray}\nonumber
        \frac{P_N^R}{\Delta
    f}=&2eV&\coth\left(\frac{eV}{2k_BT}\right)\frac{\sum_n\mathfrak{T}_n\left(1-\mathfrak{T}_n\right)}{\sum_n\mathfrak{T}_n}\\
    \label{Eq:Shot}&+&4k_BT\frac{\sum_n\mathfrak{T}_n^2}{\sum_n\mathfrak{T}_n},
\end{eqnarray}
where $\mathfrak{T}_n$ denotes the energy-independent transmission
of the $n$-th channel occupied at $T=0$ through the QPC. As
mentioned in Sec.~\ref{sec:transfer} we can assign an effective
noise temperature $T_\textrm{QPC}$ to this power spectral density
via
\begin{equation}
\frac{P_N^R}{\Delta f}=4k_BT_\textrm{QPC}.
\end{equation}
In our experiments at an ambient temperature of $2~\textrm{K}$, this
is almost entirely dominated by thermal noise, but at dilution
refrigerator temperatures, shot noise can easily exceed thermal
noise, especially if a large source-drain bias can be applied. Since
the network is reciprocal, the QPC's noise temperature will
contribute to the total setup noise proportionally to $\kappa$.

In contrast, low-frequency thermal noise in the parasitic resistor
\begin{equation}\label{Eq:Therm}
        \frac{P_N^r}{\Delta
    f}=4 k_B T_\textrm{circ}
\end{equation}
transforms via the coefficient $N$ introduced previously. In our
experiments we estimate the circuit noise temperature to be as high
as $5~\textrm{K}$, probably due to insufficient thermal anchoring of
the printed circuit board and/or heating of the matching network by
the low-temperature amplifier (due to spatial constraints there is
no isolator between resonant circuit and amplifier in our
experiments).

The frequency dependencies of the coefficients $\kappa$ and $N$ as
well as $\Gamma$ are shown in Fig.~\ref{Fig:PNnoise}(a), using our
usual circuit parameters. While matching is poor, most power is
reflected and consequently $\kappa$ and $N$ are small. Once power
starts to be transmitted, the matching network and the load share
the amount. The exact ratio between power delivered to the load an
dissipated over the lossy matching network strongly depends on the
specifics of the circuit, including the load resistance $R$. This
behavior is shown in Fig.~\ref{Fig:PNnoise}(b). We want to draw
attention to the tendency that the higher the load resistance, the
more difficult it becomes to deliver power to it, which can also be
seen from Eq.~(\ref{Eq:KSimplePozar}).

Finally the noise power of the low-noise amplifier is denoted by its
noise temperature $T_\textrm{amp}$. For a state-of-the-art low-noise
amplifier this can be as low as a few Kelvin. In our case, we use
the data sheet values of $2~\textrm{K}$ for our QuinStar U-200 unit,
and assume the amplifier's input impedance $Z_\textrm{amp}$ to be
exactly $50~\Omega$.

Thus we have all the constituents of the total system noise given by
Eq.~(\ref{Eq:SystemNoise}), and we plot its value as a function of
frequency and of load resistance at resonance in
Figs.~\ref{Fig:PNnoise}(c) and (d), respectively. For
instructiveness the contribution of the QPC has been artificially
split into pure thermal and shot noise by considering the two
limiting cases $eV\gg k_BT$ and $k_BT\gg eV$ in Eq.~(\ref{Eq:Shot}).

Whereas away from resonance amplifier noise is most relevant, the
high circuit temperature exceeds this value significantly at
resonance. At large and intermediate QPC resistances less power is
transmitted from and to the load as compared to dissipation in the
matching network, as a consequence of which noise from the QPC is
only relevant at low $R$.

\begin{figure}[htb]
        \centering
\includegraphics{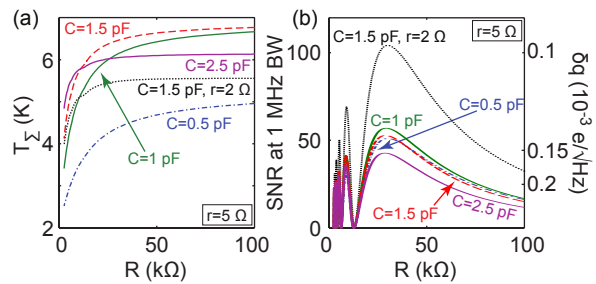}
\caption{\small (a) System noise temperature at the input of the
low-noise amplifier as a function of QPC resistance for different
series capacitances at resonance, using the same parameter values as
above. (b) Expected signal-to-noise ratio at 1 MHz bandwidth and
charge sensitivity, calculated by dividing the reflected signal
power adapted from Fig.~\ref{Fig:Vrefl}(d) by the total noise in
(a).}\label{Fig:BW}
\end{figure}

The curves of the total system noise temperature $T_\Sigma$ as a
function of load resistance for different series capacitances shown
in Fig.~\ref{Fig:BW}(a) look rather similar to each other since the
struggle between $\kappa$ and $N$ is always comparable - except for
$C=0.5$ pF (solid blue line), where matching is so poor that a
significant part of the incoming power is reflected and there is
less power transfer to and from circuit and load.

Knowing now the total noise power and the expected signal height, we
can calculate and plot the signal-to-noise ratio at a given
bandwidth ($1~\textrm{MHz}$ in Fig.~\ref{Fig:BW}(b)) and the charge
sensitivity $\delta q=e/\sqrt{\textrm{SNR}\times\textrm{BW}}$. While
for high losses in the matching network ($r=5~\Omega$) the SNR
barely depends on the series capacitance, decreasing the circuit's
losses to $r=2~\Omega$ can double it. An important point to note,
though, is that values leading to improper matching (in particular
$C=0.5~\textrm{pF}$, $r=5~\Omega$) can only compete because of the
particularly high noise contribution of the matching network. If the
total noise is limited by amplifier noise, Fig.~\ref{Fig:BW}(a) will
essentially be a set of straight lines and Fig.~\ref{Fig:BW}(b) will
look identical to Fig.~\ref{Fig:Vrefl}(d). Nevertheless, we observe
that the SNR is not extremely sensitive to the quality of the
matching as long as the circuit losses are kept at a moderate level.

It should be mentioned again that the change in reflection
coefficient $\Delta\Gamma$ was determined by modeling the QPC with a
saddle-point potential which may be inappropriate for actual QPCs in
an experiment. Thus the optimal sensitivity might also be attained
at much larger QPC resistances where their potential landscape is
often found to be more sensitive to local changes. Consequently, the
maximal SNR should rather be determined by
experiment\cite{mueller:10}.

Numerically, we accordingly expect a SNR of the order of 50 at a
bandwidth of $1~\textrm{MHz}$ given a change in QPC conductance of
$\Delta G\approx0.08\times2e^2/h$ using assumptions realistic for
our experimental setup at $2~\textrm{K}$. This corresponds to a
charge sensitivity better than
$2\times10^{-4}~e/\sqrt{\textrm{Hz}}$. We will compare these results
with experimentally obtained values in the following chapter.

\section{Measurements on an Etched QPC\label{sec:measurement}}

\begin{figure}[htb]
        \centering
\includegraphics{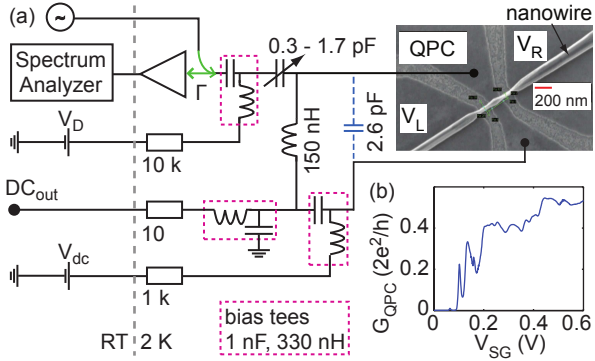}
\caption{\small (a) Schematics of the circuit used for the
measurements in this section including a scanning electron
microscopy image of the InAs nanowire quantum dot and its
self-aligned charge detector. (b) QPC conductance as a function of
side-gate voltage $V_\textrm{SG}=V_\textrm{L}+V_\textrm{R}$ at a
bias of $250~\mu$V. A total resistance of $4.26~\textrm{k}\Omega$
($2~\textrm{k}\Omega$ contact resistance and $2.26~\textrm{k}\Omega$
from the resistors on the circuit board and the room-temperature
plug as well as the dc wiring) has been
subtracted.}\label{Fig:NWsample}
\end{figure}

To demonstrate the validity of above calculations, we cross-check
them against measurements taken with a QPC etched into a shallow
two-dimensional electron gas of a GaAs/AlGaAs heterostructure (see
Fig.~\ref{Fig:NWsample}). This QPC was designed as a charge detector
for electrons trapped in a self-aligned InAs nanowire quantum
dot\cite{shorubalko:08}. The nanowire was not conductive, though,
and therefore we unfortunately could not form a quantum dot.
Consequently we will ignore the nanowire's presence henceforth.

The QPC was incorporated into the matching network as shown in
Fig.~\ref{Fig:NWsample}(a). This circuit allows for simultaneous dc
measurements and \emph{in situ} tunability through a variable
capacitance (varactor diode), and the values of the circuit
parameters correspond to the ones for which the above calculations
have been carried out (see Tab.~\ref{Tab:Values}).

The conductance through the QPC as a function of side-gate voltage
(denoted by $V_\textrm{L}$ and $V_\textrm{R}$) is depicted in
Fig.~\ref{Fig:NWsample}(b), where a resistance of
$4.26~\textrm{k}\Omega$ has been subtracted to compensate for
contact and cable resistances. The dc bias applied was
$250~\mu\textrm{V}$. Instead of the monotonous increase in
conductance from 0 to $2e^2/h$ with increasing gate voltage expected
for a QPC described by a saddle-point potential we observe several
transmission resonances and the conductance stays well below
$2e^2/h$ over the entire range of applicable gate voltages. These
resonances, presumably caused by disorder in the QPC channel, may
well influence the shot-noise behavior of our QPC.

\begin{figure}[htb]
        \centering
\includegraphics{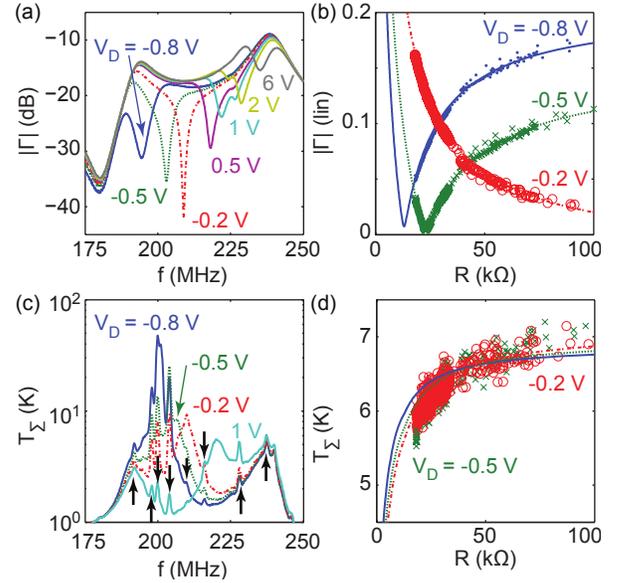}
\caption{\small (a) Absolute value of the reflection coefficient
versus frequency at zero QPC conductance for different diode
voltages. (b) As in (a), but as a function of QPC resistance at
resonance. For the calculated curves we used $L=150~\textrm{nH}$,
$C_p=2.6~\textrm{pF}$, $R_\textrm{contact}=2~\textrm{k}\Omega$,
$R_\textrm{dc wiring}=2.26~\textrm{k}\Omega$, and
$C=1.71~\textrm{pF}$, $r=5.2~\Omega$ (solid blue line),
$C=1.60~\textrm{pF}$, $r=4.55~\Omega$ (dotted green line), or
$C=1.39$ pF, $r=4.85~\Omega$ (dashed red line). The measured
reflection was converted to linear scale using an overall
attenuation of $20~\textrm{dB}$ (blue dots), $17.8~\textrm{dB}$
(green crosses), and $17.8~\textrm{dB}$ (red circles). (c) Measured
total noise in the absence of a carrier signal. A frequency
independent gain of $45.8~\textrm{dB}$ was subtracted from the
signal measured with a spectrum analyzer at a resolution bandwidth
of $1~\textrm{MHz}$ and a video bandwidth of $10~\textrm{Hz}$. The
arrows mark spurious noise peaks occurring in the spectrum. (d)
System noise temperature as a function of QPC resistance, measured
with a resolution and video bandwidth of $100~\textrm{kHz}$ and
$10~\textrm{Hz}$, respectively. The measured curves have been
brought to the same level by subtracting a gain of $46~\textrm{dB}$
for the green xs ($V_\textrm{D}=-0.5~\textrm{V}$) and
$44.8~\textrm{dB}$ for the red circles
($V_\textrm{D}=-0.2~\textrm{V}$). For the calculations we used the
same parameters as in (b) with the addition of
$T_\textrm{amp}=1.8~\textrm{K}$, $T_\textrm{QPC}=2~\textrm{K}$,
$T_\textrm{circ}=5.3~\textrm{K}$,
$V_\textrm{bias}=250~\mu\textrm{V}$, and we assumed a perfectly
matched amplifier input.}\label{Fig:NWnoise}
\end{figure}

In Fig.~\ref{Fig:NWnoise}(a) we demonstrate the tunability of the
resonance frequency by applying a reverse-bias voltage to the
varactor diode. We clearly see an increase in resonance frequency
for larger diode voltages (smaller diode capacitances) and observe a
tunability range from roughly 190 to $230~\textrm{MHz}$. The
background of these curves is set by the frequency-dependent gain of
the cryogenic low-noise amplifier in combination with standing waves
building up between the matching network and the amplifier input.
The magnitude of these standing waves is in agreement with a quoted
input standing wave ratio of 1:2.5 of the cryogenic amplifier. Here,
the QPC was pinched off completely.

Measuring the reflection at resonance as a function of QPC
resistance yields Fig.~\ref{Fig:NWnoise}(b). Again, a contact and
wiring resistance of $4.26~\textrm{k}\Omega$ has been taken into
account. Furthermore, an overall attenuation of $20~\textrm{dB}$
($V_\textrm{D}=-0.8~\textrm{V}$; blue dots), $17.8~\textrm{dB}$
($V_\textrm{D}=-0.5~\textrm{V}$; green crosses), and
$17.8~\textrm{dB}$ ($V_\textrm{D}=-0.2~\textrm{V}$; red circles) has
been subtracted from the measured $\Gamma$ to atone for the
attenuators, the directional coupler, losses in the rf lines,
standing waves, and the gain of the amplifier, whereafter the result
was converted to a linear (voltage) scale. The solid blue, dotted
green, and dashed red lines are calculations of
$\Gamma=(Z-Z_0)/(Z+Z_0)$ at resonance, using $L=150~\textrm{nH}$,
$C_p=2.6~\textrm{pF}$, $R_\textrm{contact}=2~\textrm{k}\Omega$,
$R_\textrm{dc wiring}=2.26~\textrm{k}\Omega$, and
$C=1.71~\textrm{pF}$, $r=5.2~\Omega$
($V_\textrm{D}=-0.8~\textrm{V}$; solid blue line),
$C=1.60~\textrm{pF}$, $r=4.55~\Omega$
($V_\textrm{D}=-0.5~\textrm{V}$; dotted green line), and
$C=1.39~\textrm{pF}$, $r=4.85~\Omega$
($V_\textrm{D}=-0.2~\textrm{V}$; dashed red line). We note that the
difference in parasitic losses $r$ are rather large and
non-monotonic, which might stem from a frequency dependence of $r$
due to a frequency-dependent fraction of rf current passing the
inductance and capacitance. On the other hand, keeping $r$ equal for
all three calculated curves would move the matched resistances
further apart.

Though there are in principle 11 independent parameters in the
calculations for Fig.~\ref{Fig:NWnoise}(b), knowing $L$ and having
measured the three resonance frequencies ($198.26~\textrm{MHz}$;
$205.55~\textrm{MHz}$; $210.74~\textrm{MHz}$) fixes the sum of
$C_p+C_i$. From experience and from gauging diode capacitances at
low temperatures we have a rather good idea of the values of each of
those capacitances. Using these guesses in combination with the
position of the minima in $\Gamma$ we can determine the $r_i$.
Finally, the attenuations are set via the value of $\Gamma$ for
large QPC resistance, and as a \emph{a posteriori} justification we
can compare the results with the background of
Fig.~\ref{Fig:NWnoise}(a). Thus we feel confident about the
appropriateness of the extracted circuit-parameter values.

Figure \ref{Fig:NWnoise}(c) shows the frequency dependence of the
total noise in our experimental setup measured with a spectrum
analyzer at a resolution bandwidth of $1~\textrm{MHz}$ and video
averaging at $10~\textrm{Hz}$. No rf signal was applied, and the QPC
was held at a constant resistance of roughly $120~\textrm{k}\Omega$.
We converted the measured power spectral density to a noise
temperature by subtraction of a frequency-independent gain of
$45.8~\textrm{dB}$ and division by the Boltzmann constant.

We observe an unexpectedly high frequency dependence of the noise
temperature, accented by spurious noise peaks - marked by black
arrows - of unknown origin\cite{Note5}, changing their magnitude
when the diode capacitance is altered. Values of up to 50 K are
visible at $200~\textrm{MHz}$ while the noise temperature is well
below $10~\textrm{K}$ above $215~\textrm{MHz}$, which necessitates
an according choice of resonance frequency when trying to detect
single-electron charging. Furthermore, we can see the standing-wave
pattern, revealing at which frequencies noise from the QPC and the
matching network can couple into the amplifier.

On resonance, the noise level is raised by a few degrees (not
regarding the spurious peaks) as expected due to our elevated
ambient temperature. A more detailed view of the dependence of the
system noise temperature on the matching network is given in
Fig.~\ref{Fig:NWnoise}(d), where the total noise is plotted as a
function of QPC resistance. Here, the resolution bandwidth was
reduced to $100~\textrm{kHz}$, while the video bandwidth was kept at
$10~\textrm{Hz}$. By subtracting a gain of $46~\textrm{dB}$ (green
xs, $V_\textrm{D}=-0.5~\textrm{V}$) and $44.8~\textrm{dB}$ (red
circles; $V_\textrm{D}=-0.2~\textrm{V}$) the measured curves have
been brought to the same level to negate the effect of the spurious
noise peaks on the performance at $V_\textrm{D}=-0.5~\textrm{V}$.
The calculated curves use the same parameter values as above,
including $T_\textrm{amp}=1.8~\textrm{K}$,
$T_\textrm{QPC}=2~\textrm{K}$, $T_\textrm{circ}=5.3~\textrm{K}$, and
a dc bias $V_\textrm{bias}=250~\mu\textrm{V}$. We also assume a
perfectly matched amplifier input.

The circuit temperature of $5.3~\textrm{K}$ seems somewhat large,
but may be explained by poor thermal anchoring of the matching
network inside the vacuum cap of the low-temperature stage of our
experimental insert.

The system noise is largest for highest load resistances and
decreases for lower $R$ as power transfer from the matching circuit
to the low-noise amplifier - quantified by $N$ - is suppressed. We
also realize that the exact values of $r$ and $C$ are not extremely
important for the overall system noise (within a certain window), as
expected from our considerations in the previous section. Also, our
model's predictions follow the experimental curves rather well, and
the subtracted gains are quite plausible.

\begin{figure}[htb]
        \centering
\includegraphics{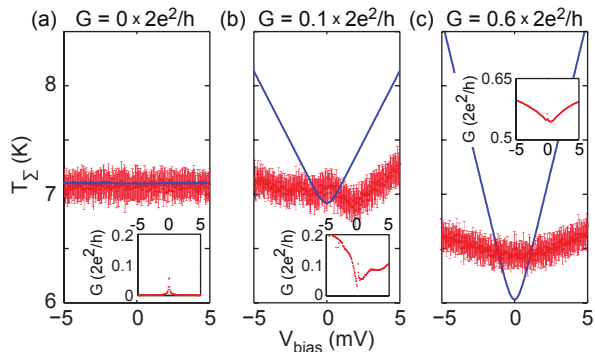}
\caption{\small Total noise temperature at resonance in the absence
of a rf carrier signal for $V_\textrm{D}=-0.8$ V as a function of dc
bias for different QPC conductances. The spectrum analyzer's
resolution bandwidth was set to 100 kHz with a video bandwidth of 10
Hz. We subtracted a constant gain of 45.1 dB. (a)
$G_\textrm{QPC}=0\times2e^2/h$. (b)
$G_\textrm{QPC}=0.1\times2e^2/h$. (c)
$G_\textrm{QPC}=0.6\times2e^2/h$. The common parameters used for the
calculation are identical to the ones used for
Fig.~\ref{Fig:NWnoise}(d). Insets: Conductance of the QPC as a
function of applied bias.}\label{Fig:NWshot}
\end{figure}

Shot-noise calibration is a very powerful tool for determining
system noise temperatures since the noise power of a nanostructure
as a function of dc bias can generally be known very accurately.
Therefore, we recorded the noise power with a resolution and video
bandwidth of $100~\textrm{kHz}$ and $10~\textrm{Hz}$, respectively,
as a function of dc bias and show the result for different QPC
conductances in Fig.~\ref{Fig:NWshot}. The voltage on the varactor
diode was $-0.8~\textrm{V}$, and we subtracted a gain of
$45.1~\textrm{dB}$. The conductance values chosen were
$G_\textrm{QPC}=0\times2e^2/h$ (a), $G_\textrm{QPC}=0.1\times2e^2/h$
(b), and $G_\textrm{QPC}=0.6\times2e^2/h$ (c) determined by the
current at $+5~\textrm{mV}$ dc bias (see insets) and taking into
account contact and wiring resistances. The error bars denote the
statistical spread of repeated measurements.

As expected we do not observe any increase in noise for increasing
bias when the QPC is completely pinched off ($\kappa\approx0$). For
non-zero conductance, though, the bias dependence is much weaker
than predicted (with the same parameters as above) and for
$G_\textrm{QPC}=0.1\times2e^2/h$, the noise is even increasing
non-monotonically. Similarly, the dc conductances shown in the
insets are not constant. While at bias voltages of several mV we are
already well in the non-linear regime, the predictions of
Eq.~(\ref{Eq:Shot}) should still hold qualitatively.

A reason why we observe too small values for the shot noise is
partially the reflection of incoming signals at the cryogenic
amplifier's input. Roughly $40\%$ of the shot noise may thus be
reflected, which, however, still does not fully explain the low
observed noise levels at high bias. A further explanation may be an
overestimation of power transfer $\kappa$ between QPC and amplifier.
Additionally, localized states in the QPC leading to the atypical
conductance curves in the insets of Fig.~\ref{Fig:NWshot} and in
Fig.~\ref{Fig:NWsample}(b) may strongly influence the shot-noise
behavior of the QPC. In particular, it may be conceivable that shot
noise is suppressed even at non-integer conductance values since the
transmission through the QPC reaches a plateau well before $2e^2/h$.

Also, our predicted noise temperatures do not exactly match the
measured noise at zero bias for high conductance. This most probably
stems from the height of the spurious noise peaks in
Fig.~\ref{Fig:NWnoise}(c) which is found to depend on $R$.

To summarize this experimental section, we can model the system
noise in our setup very well with the exception of the unexpected
magnitude and shape of the shot-noise curves. Potentially it would
be easier to accurately determine the shot noise with a smaller
background noise.

With a similar device as the one shown in
Fig.~\ref{Fig:NWsample}(a), where the nanowire quantum dot did form,
we have obtained a charge sensitivity of approximately
$4\times10^{-4}~e/\sqrt{\textrm{Hz}}$, slightly worse than our
estimate in Sec.~\ref{sec:noise}. The difference can be attributed
to signal loss from standing waves and slightly higher noise in the
experimental setup - mainly the matching circuit.

\section{Conclusion\label{sec:outro}}
In a generally valid derivation for reciprocal two-port networks we
have demonstrated a charge-detection experiment's signal-to-noise
ratio to increase linearly with power transfer to the load. Thereby
we have found it important that losses in the resonant circuit are
minimized, unless shot noise is strongly dominating the overall
noise.

Subsequently we have discussed \emph{in situ} tunability in great
detail and have visualized the effect of changing the series
capacitance of a $C-LCR$ circuit. We have also calculated the
maximally allowed input voltage and power when limiting the voltage
drop over the load, and we have thereby estimated the signal height
of the matching network's response to a change in the charge state
of the system to amount to several hundred nV.

By assessing the total noise in the system we could determine the
expected overall SNR and charge sensitivity which we found to be
only slightly influenced by the exact circuit-parameter values as
long as a fair quality of matching is achieved. This fact is
substantiated by previous experiments\cite{mueller:10}. The values
for the charge sensitivity one can expect in a similar experiment
range in a few $10^{-4}~e/\sqrt{\textrm{Hz}}$.

With our present experiments we can confirm tunability and the
circuit model as well as the noise levels in the system. We did
observe spurious noise peaks in the spectrum, though, which were
depending on the circuit impedance. We attributed them to the
low-noise amplifier since they occurred also for zero applied rf or
dc bias. Furthermore the amount of shot noise we detected was
significantly lower than expected. This could partially be explained
by the fact that a mismatched amplifier input reflects the noise
signal back (on the other hand, this would require higher circuit
noise temperatures), and maybe localizations in the QPC - revealing
themselves by an unusual gate-voltage behavior and nonlinear $I-V$
characteristics of the QPC - cause the shot noise to be suppressed.

While we still recommend the point of maximal sensitivity to be
determined by experiment, our considerations should give an
experimentalist dealing with a similar circuit a good feeling for
the mutual influences of the circuit elements which is of great
importance for optimizing the measurement parameters. Also, in our
discussions on the power transfer of noise we identified the
parameter relevant for improvement in different regimes.

Since a good power transfer to the charge detector is quite
important, it would be very helpful indeed to be able to determine
the source of the losses in such a circuit more closely. We believe
that such a study would require a detailed analysis  - comparable to
Ref.~\onlinecite{hellmueller:12} - of the influence of the exact
sample design (e.g. bonding wires, ohmic contacts, distance traveled
in the two-dimensional electron gas) on the matching network's
losses.

\section*{Acknowledgements}
We are greatly indebted to Cecil Barengo and Paul Studerus for their
superb technical assistance, and funding by the Swiss National
Science Foundation (SNF) via National Center of Competence in
Research (NCCR) Nanoscale Science is gratefully acknowledged.


\begin{thebibliography}{30}
\bibitem{NoteTemp}
  The sign of $T_\textrm{QPC}-T_\textrm{circ}$ determines whether
the SNR as a function of $\kappa$ is straight, concave or convex.
\end{thebibliography}

\end{document}